\documentclass[12pt,preprint]{aastex}
\usepackage{psfig}
\newcommand{\lta}{$\; \buildrel < \over \sim \;$}
\newcommand{\simlt}{\lower.5ex\hbox{\lta}}
\newcommand{\gta}{$\; \buildrel > \over \sim \;$}
\newcommand{\simgt}{\lower.5ex\hbox{\gta}}

\newcommand{\cm}{{\rm\,cm}}
\newcommand{\kms}{{\rm\,km\,s^{-1}}}
\newcommand{\msun}{{\,M_\odot}}

\newcommand{\pb}{\phi_{\scriptscriptstyle 98}}
\newcommand{\ps}{\phi_{\scriptscriptstyle 67}}

\newcommand{\mwd}{{\,M_{\rm WD}}}
\newcommand{\rwd}{{\,R_{\rm WD}}}
\newcommand{\w}{{\omega}}
\newcommand{\W}{{\Omega}}
\newcommand{\wW}{{\w-\Omega}}
\newcommand{\wWW}{{\omega-2\Omega}}
\newcommand{\ha}{{{\rm H}\alpha}}
\newcommand{\hb}{{{\rm H}\beta}}
\newcommand{\hg}{{{\rm H}\gamma}}

\slugcomment{Accepted to AJ}
\shortauthors{Belle et al.}
\shorttitle{X-ray and Optical Observations of EX Hya}

\begin{document}

\title{Simultaneous X-ray and Optical Observations of EX Hydrae
\footnote{Based in part on observations obtained with the Apache Point
Observatory 3.5-meter telescope, which is owned and operated by the
Astrophysical Research Consortium.} }

\author{K. E. Belle\altaffilmark{2}, 
S. B. Howell\altaffilmark{3}, K. Mukai\altaffilmark{4,5}, 
P. Szkody\altaffilmark{6}, 
K. Nishikida\altaffilmark{7},
D. R. Ciardi\altaffilmark{8},
R. E. Fried\altaffilmark{9},
J. P. Oliver\altaffilmark{10}
}

\altaffiltext{2}{X-2 MS B227, Los Alamos National Laboratory, Los Alamos,
NM 87545; belle@lanl.gov}
\altaffiltext{3}{WIYN Observatory \& NOAO, P.O. Box 26732,
950 N. Cherry Ave., Tucson, AZ 85726-6732; howell@noao.edu}
\altaffiltext{4}{NASA GSFC, Code 662  Lab. for High Energy Astrophysics,  
Greenbelt, MD 20771; mukai@milkyway.gsfc.nasa.gov}
\altaffiltext{5}{Also Universities Space Research Association}
\altaffiltext{6}{University of Washington, Dept. of Astronomy,  Box 351580 
Seattle, WA 98195; szkody@astro.washington.edu}
\altaffiltext{7}{University of California, Berkeley, Space Sciences Lab,
Berkeley, CA 94720; kaori@ssl.berkeley.edu}
\altaffiltext{8}{Michelson Science Center/Caltech, 770 S. Wilson Ave.,
Mail Code: 100-22, Pasadena, CA 91125; ciardi@ipac.caltech.edu}
\altaffiltext{9}{Braeside Obs., P.O. Box 906   Flagstaff, AZ 86002-0906}
\altaffiltext{10}{University of Florida, Dept. of Astronomy, PO Box 112055
Gainesville, FL 32611; oliver@astro.ufl.edu}

\begin{abstract}
The intermediate polar, EX Hydrae, was the object of a large
simultaneous multiwavelength observational campaign during 2000 May --
June.  Here we present the Rossi X-ray Timing Explorer photometry and
optical photometry and spectroscopy from ground-based observatories
obtained as part of this campaign.  Balmer line radial velocities and
Doppler maps provide evidence for an extended bulge along the outer
edge of the accretion disk and some form of extended/overflowing
material originating from the hot spot.  In addition, the optical
binary eclipse possesses an extended egress shoulder, an indication
that an additional source (other than the white dwarf) is coming out
of eclipse.  We also compare the X-ray and optical results with the
results obtained from the EUV and UV observations from the
multiwavelength data set.
\end{abstract}


\section{INTRODUCTION}\label{xray_intro}
Intermediate polars (IPs) are a class of magnetic cataclysmic variable
(CV) in which an asynchronously rotating white dwarf accretes matter
from an accretion disk via magnetically controlled accretion curtains.
Modulations observed over the white dwarf spin period in IPs are
generally attributed to either photoelectric absorption by the
accretion curtain \citep[the accretion curtain
model,][]{cord85,rosen88} or to a self-eclipse of the lower accretion
pole by the white dwarf \citep[e.g.,][]{beuer88}.  Over the binary
orbital period, the observed modulation is due to the properties of a
typical CV (accretion disk, eclipses by the secondary) and also
eclipses of the emitting regions close to the white dwarf surface.

Here we report on simultaneous X-ray and optical photometric and
optical spectroscopic observations of the IP EX Hydrae.  EX Hya has a
spin period of 67 minutes (the spin phase is denoted as $\ps$
throughout) and a binary orbital period of 98 minutes (binary phase is
denoted as $\pb$ throughout). The mass of the white dwarf has yet to
be constrained completely, as optical and UV studies give
$\mwd\sim0.8\msun$ \citep[e.g.,][]{hel87,bel03}, while X-ray studies
give $\mwd\sim0.5\msun$ \citep[e.g.,][]{fuj97,hoo04}.  Reports on
recent X-ray spectroscopic observations of EX Hya can be found in
\citet{mau01}, \citet{mau03}, \citet{muk03}, and \citet{hoo04}.
\citet{mau01} and \citet{mau03} find that the hot, $T_{\rm e}\sim 12$
MK, emitting plasma in the accretion column has an electron density of
$n_{\rm e}=1.0^{+2.0}_{-0.5} \times 10^{14}\cm^{-3}$, using Chandra
High Energy Transmission Grating (HETG) spectra.  \citet{muk03} find
that the Chandra HETG spectrum of EX Hya can be fit with a simple
isobaric cooling flow model.  \citet{hoo04} use the Chandra spectrum
to find a radial velocity and mass for the white dwarf,
$K_1=58.2\pm3.7\kms$ and $\mwd=0.49\pm0.13\msun$.

This paper is the third and final paper in a series detailing
simultaneous multiwavelength observations of EX Hya.  In the first
paper \citep{bel02}, we reported on observations with the {\it Extreme
Ultraviolet Explorer} ({\it EUVE}); its one million seconds of
photometry and spectroscopy of EX Hya provided the basis for the
multiwavelength observations.  The EUV photometry revealed the
presence of two dips in the binary light curve, whose absorption
depths changed with spin phase.  Our second paper \citep{bel03}
presented {\it Hubble Space Telescope} ({\it HST}) spectroscopy of EX
Hya.  The mass of the white dwarf was calculated as
$\mwd=0.91\pm0.05\msun$ from the $K$ amplitude of the radial velocity
curve of the narrow UV emission lines.  Spectral model fits to the UV
data also produced the same mass white dwarf with $T=23,000$ K and an
accretion disk truncated at $2.5\rwd$.

\section{OBSERVATIONS}
As part of a simultaneous multiwavelength observational campaign of EX
Hya \citep[executed in 2000 May -- June, see][]{bel02,bel03}, we
obtained X-ray photometry from the Rossi X-ray Timing Explorer
satellite ({\it RXTE}), optical spectroscopy from Apache Point
Observatory (APO) and optical photometry from Braeside Observatory
(BO) and Rosemary Hill Observatory (RHO).  Table~\ref{obslog} gives a
complete list of the simultaneous EX Hya observations completed during
2000 May -- June; those marked with an asterisk are presented in this
paper.

{\it RXTE} observed EX Hya for 15 ks over four separate observations
using the Proportional Counter Array, which consists of five
Proportional Counter Units (PCUs).  During the observations, only PCU2
and PCU3 were in continuous operation, so we present here only the
data from these two PCUs.  The data set spans the wavelength range
$1-5$ \AA\ ($12.4-2.5$ keV) and provides complete coverage of the spin
period but only partial coverage of the binary period, lacking binary
phases $\pb=0.30-0.39$.  Sixteen nights of $BVR$ photometry were
obtained from BO and 10 nights of $I$ photometry were obtained from
RHO.  Complete binary and spin phase coverage was obtained with all of
the optical photometry.

Blue and red spectra were obtained during two nights of observations,
2000 May 15 and 28 (UT), with the 3.5 m telescope at APO, which
provided complete coverage of the binary and spin phases of EX Hya.
The observations utilized the Double Imaging Spectrograph and obtained
high resolution ($0.80$ \AA\ pixel$^{-1}$) blue spectra in the
wavelength range $4185-5010$ \AA\ and high resolution ($1.17$ \AA\
pixel$^{-1}$) red spectra in the wavelength range $6285-7320$ \AA.
These wavelength ranges include the $\hg$ $\lambda4340.5$\AA, $\hb$
$\lambda4861.3$\AA, and $\ha$ $\lambda6562.8$\AA\ emission lines.

\section{ANALYSIS}
\subsection{X-ray and Optical Photometry}
\subsubsection{Period Analysis}
We first searched for periods in each of our photometric data sets
using the phase dispersion minimization (PDM) routine \citep{stell78},
a technique for finding periods in a data set by minimizing scatter
about a light curve.  For a given input data set and range of test
frequencies, PDM will compute a light curve and a binned light curve
from the input data for each test frequency, and the dispersion of the
data about the mean light curve.  The dispersion of the data will be
at a minimum when a 'real' frequency is found.  This is reflected in
the value of $\Theta$ that PDM calculates.  A two-sided F-test can
then be used to calculate the confidence of a given frequency for
which $\Theta$ is at a minimum.

\begin{figure}
\plotone{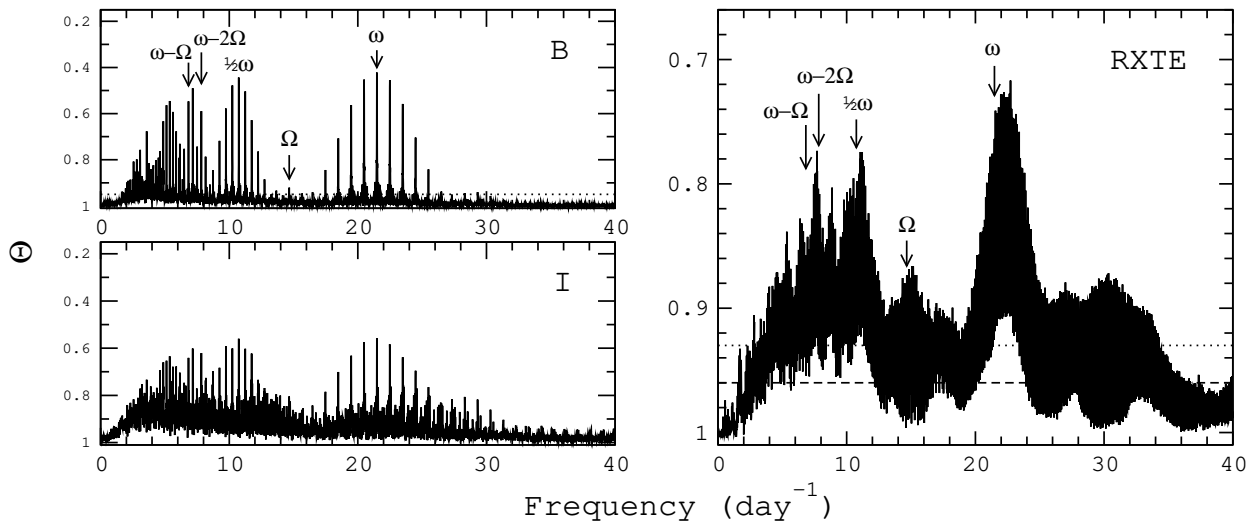}
\caption{Thetagrams for the $B$, $I$, and {\it RXTE} photometry.  A
smaller $\Theta$ value indicates a higher statistical confidence (note
the y-axis has been inverted).  Frequencies commonly found in IPs are
labeled on the $B$ and {\it RXTE} plots and listed in Table~\ref{freq}.
The dotted lines represent the 99\% confidence level, while the dashed
line on the {\it RXTE} plot represents the 95\% confidence level.  The
99\% confidence level for the $I$ thetagram is at 0.99.  Values of
$\Theta$ for the frequencies noted in this figure are given in
Table~\ref{theta}.\label{bix_theta}}
\end{figure}

The PDM results are shown as thetagrams in Figure~\ref{bix_theta} for
the $B$, $I$, and {\it RXTE} photometry.  Thetagrams were also created
for the $V$ and $R$ photometric data, but are similar to the $B$
thetagram (the same frequencies are found), so are not presented here.
Strong frequencies in the $B$ and $I$ thetagrams, above the 95\%
confidence level, include the spin frequency, $\w$, the binary
frequency, $\W$, harmonics of these two frequencies, $\w/2$ and $2\W$,
and the beat frequencies, $\wW$ and $\w-2\W$.  They have been labeled
on the {\it RXTE} and $B$ plots in Figure~\ref{bix_theta} and are
listed in Table~\ref{freq}.  The 99\% confidence levels have been
denoted on the plots as dotted lines.  The frequencies shortward of
$\sim6$ day$^{-1}$ represent the length of the observations and also
the time between data sets.

The strongest frequency in each thetagram is $\w$, with $\Theta=0.76$
($RXTE$), 0.42 ($B$), and 0.56 ($I$).  Though several other
frequencies appear above the 95\% confidence level in the $RXTE$
thetagram, the time-limited nature of the photometric data precludes
us from drawing many conclusions about the observed frequencies.  The
most we can say is that the possible appearance of the beat
frequencies $\wW$ and $\wWW$ may be interesting, but more data is
required to confirm their existence.  We would also like to note that
during the {\it RXTE} observation, the satellite orbital period was
$95.73$ min, which is very close to the $98.26$ min orbital period of
EX Hya.  The satellite orbital period would contribute to a peak in
frequency near the binary orbital period in the {\it RXTE} thetagram.
$\Theta$ values for statistically significant frequencies are given in
Table~\ref{theta}.

Previous period analyses of X-ray photometry of EX Hya are limited.  A
search of the literature found only one instance: \citet{allan98}
produced a power spectrum of their 40 ks ASCA observation ($0.6-10.0$
keV), which shows peaks at $\w$ and $\W$.  Optical thetagrams are a
bit more common and a nice example can be found in \citet{siegel89},
who report the appearance of the frequencies $\w$, $\W$, $\W/2$, and
$\w\pm2\W$.  The appearance of $\wW$ is questionable in the
\citet{siegel89} data set, and $\w+\W$ is not seen.  While $\w+\W$ is
also not seen in our data sets, $\wW$ is prominent.  Emission at the
beat frequency $\wW$ is due to reprocessed light from the secondary
and hot spot along with orbital absorption of emission from the white
dwarf and reprocessed light from the inner accretion disk and
accretion stream.  The appearance of this frequency in our data set
suggests that something has changed within the binary system that
would contribute to emission and absorption at the $\wW$ beat
frequency, which is most likely an extended hot spot region along the
outer edge of the disk.  We will explore the appearance of the $\wW$
beat frequency further in \S\ref{swave}.

\subsubsection{RXTE Photometry}
\begin{figure}
\plotone{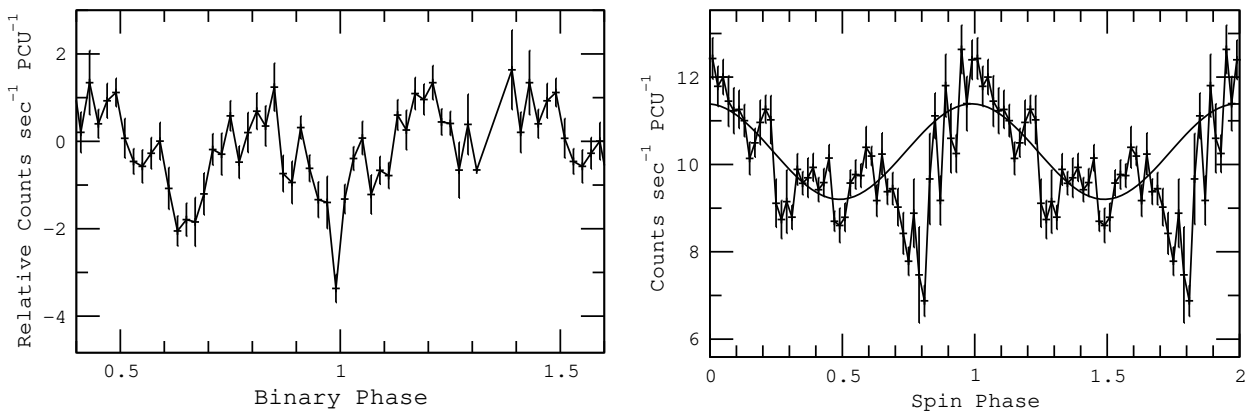}
\caption{{\it RXTE} photometry phased on the binary and spin
ephemerides.  The left panel shows binary phased data after
subtraction of the spin modulation.  The solution for the sine curve
shown fit to the spin phased data in the right panel is given in Table
\ref{sp_fit}.
\label{x_phot}}
\end{figure}

Photometric data obtained from {\it RXTE} are shown in
Figure~\ref{x_phot} phased on the binary ephemeris, $T=2437699.94179 +
0.068233846(4)\mathrm E$, and spin ephemeris, $T=2437699.8914(5) +
0.046546504(9)\mathrm E - 7.9(4)\times 10^{-13} \mathrm E^2$, (binned
to 0.02 in phase) of \citet{ephem}.  The binary phased light curve,
shown in the left panel of Figure~\ref{x_phot}, exhibits an eclipse at
$\pb\sim0.0$ of the X-ray emitting region near the white dwarf
surface.\footnote{The sinusoidal form of the $O-C$ values given in
\citet{ephem} gives $O-C=-0.005$ P$_{\rm orb}$ for the 2000 May --
June observations.  Given the moderate time resolution of our data as
presented, this shift in phase will be unnoticeable.}  In addition to
the binary eclipse, the light curve also displays quite a bit of
variability throughout its entirety.  This is a result of the combined
effects of the X-ray flickering behavior of EX Hya and the poor phase
coverage of the {\it RXTE} data.  Each phase bin was covered only once
or twice during the entire observation, in which case any flickering
in individual cycles of EX Hya will not be averaged out.  In
particular, the flickering seen at $\pb\sim0.65$ should not be
confused with the bulge eclipse at $\pb\sim0.7$ seen in lower energy
X-ray data \citep[$<2$ keV, e.g.,][]{cord85,rosen88}.

The binary phased light curve has had the sinusoidal spin modulation
subtracted.  On an absolute count rate scale, it can be seen that the
binary eclipse does not go to zero, implying a partial eclipse of the
X-ray emitting regions.  This is a result of the lower accreting pole
being eclipsed by the secondary, while the upper pole remains in view.
Unfortunately, our $RXTE$ data set is not extensive enough to analyze
binary light curves extracted at different phase segments of the spin
period.

In creating the spin phased light curve, shown in the right panel of
Figure~\ref{x_phot}, we omitted data from binary phases
$\pb=0.95-1.01$, which correspond to the binary eclipse.  The X-ray
flickering present in the binary phased light curve is also seen
throughout the spin phased light curve.  Ignoring this flickering, one
can see that the data folded over spin phase exhibit a roughly
sinusoidal modulation, peaking at $\ps\sim0$, which is the signature
of a rotating white dwarf accreting at both magnetic poles.  The
sinusoidal fit to the spin phased data was created omitting the dip at
$\ps\sim0.8$; the solution of the form $A+B\sin(\ps-\phi_0)$ is given
in Table~\ref{sp_fit}.

\subsubsection{Optical Photometry}
\begin{figure}
\plotone{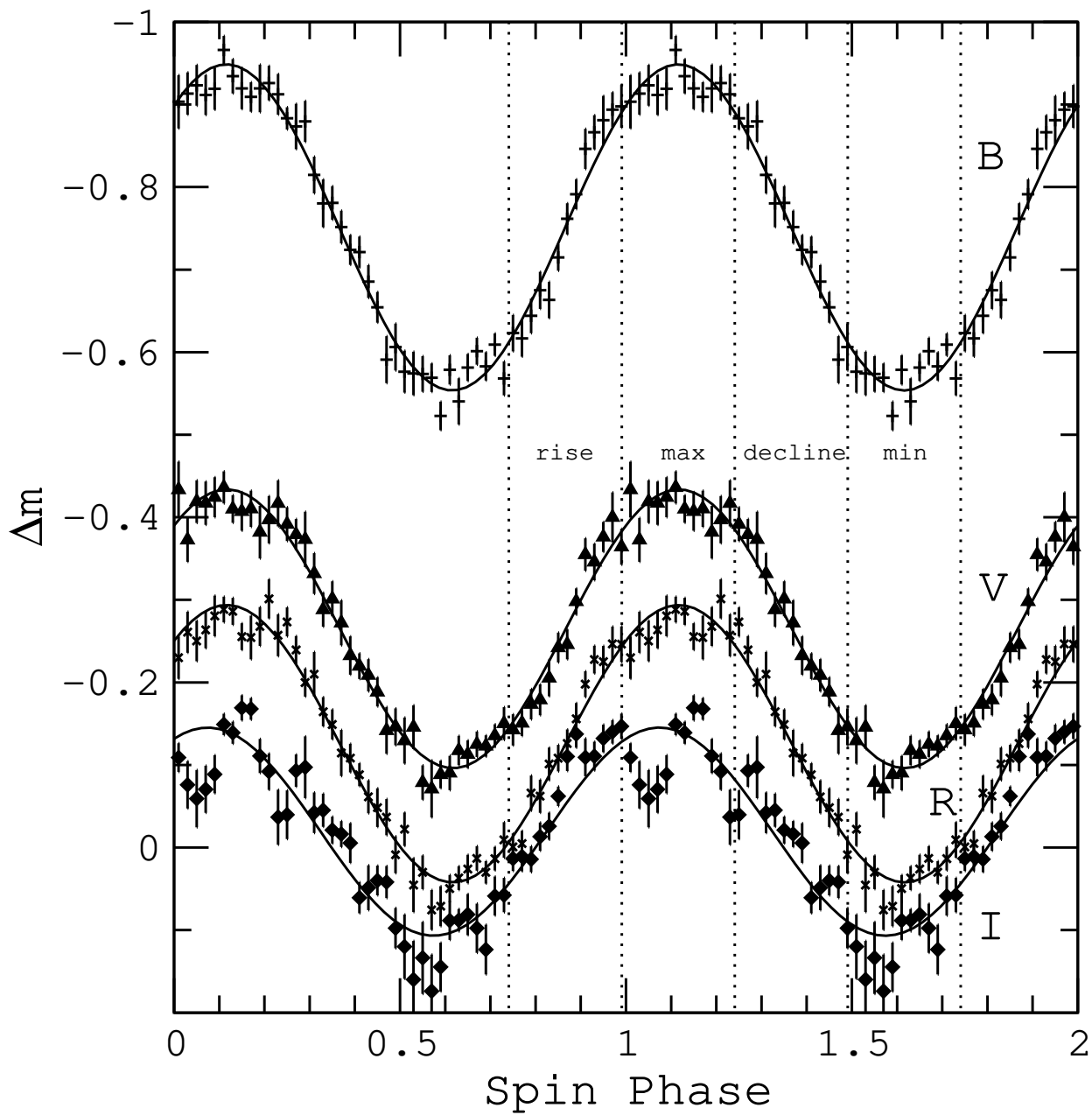}
\caption{$BVRI$ optical photometry folded on the spin period.  A
sinusoidal function is shown fit to each light curve; solutions are
given in Table \ref{sp_fit}. Phase delineations are given for spin
maximum ($\ps=0.99-1.24$), minimum ($\ps=0.49-0.74$), rise
($\ps=0.74-0.99$), and decline ($\ps=0.24-0.49$).  The offsets between
the light curves are real magnitude differences.
\label{sp_lc}}
\end{figure}

Figure~\ref{sp_lc} displays the optical photometry of EX Hya folded on
the spin ephemeris and binned to 0.02 in phase.  Marked on the Figure
are the spin phase designations of maximum ($\ps=0.99-1.24$), minimum
($\ps=0.49-0.74$), rise ($\ps=0.74-0.99$), and decline
($\ps=0.24-0.49$), which will be used for the analysis of the
photometry folded on the binary orbital period.  Data from the binary
eclipse, $\pb=0.95-1.01$, have been omitted in the spin phased light
curves.  The $BVR$ light curves are modulated sinusoidally (solutions
for the sinusoidal fits of the form $A+B\sin2\pi(\ps-\phi_0)$ are
given in Table \ref{sp_fit}) and peak at $\ps=0.12\pm0.02$.  The $I$
light curve contains quite a bit of ``wiggle'', which is a result of
modulations over the binary orbital period translating into the spin
phasing.  We do not believe that any of the $I$ modulation is due to
the secondary star.  The optical spectra (out to $7200$\AA) as well as
an infrared spectrum obtained by one of us (SBH) during the campaign
do not reveal any spectral features that could be associated with the
secondary.

The maximum phase of $\ps=0.12$ of the spin phased light curves
matches well with the value of $\ps=0.115\pm0.001$ determined from our
{\it EUVE} photometry \citep{bel02}, but differs from the value of
$\ps\approx0.0$ obtained using the UV continuum flux from our {\it
HST} data \citep{bel03}, and also the value of $\ps\sim0$ we determine
in this paper for the X-ray photometry.  The discrepancy between the
phases of spin maximum will be addressed in $\S$\ref{compare}.

\begin{figure}
\plotone{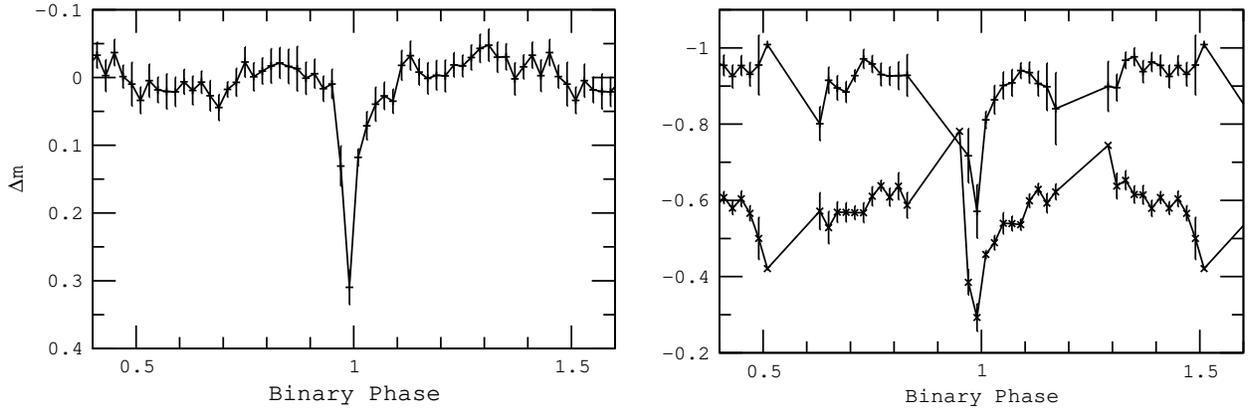}
\caption{$B$ photometric data phased on the binary period.  The
left panel displays the $B$ photometry after subtraction of the
sinusoidal spin modulation.  The right panel displays the same
photometric data separated into spin maximum (top) and spin minimum
(bottom). \label{bp_lc}}
\end{figure}

Figure~\ref{bp_lc} displays representative light curves of the optical
photometry folded on the binary ephemeris.  The left panel of
Figure~\ref{bp_lc} shows the $B$ binary phased light curve after
the sinusoidal spin modulation has been subtracted. The right panel of
Figure~\ref{bp_lc} shows the $B$ data separated into spin maximum
(top) and spin minimum (bottom).  Each light curve has been binned to
0.02 in phase.  The light curves exhibit typical behavior of an
eclipsing binary system, with an eclipse near phase $\pb=0.0$.  This
eclipse has typically been associated with an occultation of (part of)
the white dwarf by the secondary star.  Recently, though, it was
questioned if this was a white dwarf eclipse, as the binary eclipse in
EUV data of EX Hya appears at $\pb=0.97$ \citep{bel02}; it was
suggested that the optical eclipse was that of the hot spot on the
outer edge of the accretion disk.  Eclipse timings by \citet{siegel89}
show that the eclipse minimum shifts between spin phases $\ps\sim0.25$
and $\ps\sim0.75$ by about 20 s.  Assuming a binary separation of
$a\simeq5\times10^{10}\cm$, they determine that the shift in eclipse
minimum corresponds to an eclipse of material located (roughly) on the
white dwarf surface.  While our data do not have the high time
resolution required to perform a detailed analysis such as this, close
inspection of the unbinned light curves shows that the binary eclipse
during $\ps\sim0.75$ (rise) occurs ahead of the binary eclipse during
$\ps\sim0.25$ (decline) by $\sim20$ s, consistent with the earlier
results of \citet{siegel89}.

The spin separated light curves shown in the right panel of
Figure~\ref{bp_lc} reveal additional information about the binary
system.  Inspection of the binary eclipse shows an extended egress
shoulder on the spin minimum eclipse.  This behavior is also seen in
$V$, $R$, and $I$.  This shoulder would be caused by an extended
object (as compared with the white dwarf) coming out of eclipse, such
as the bulge along the outer edge of the accretion disk, or, as we
will discuss later, a region at the inner accretion disk radius where
an overflow stream impacts with the magnetosphere (such a geometry
would be favored during spin minimum).  A similar feature in the
binary eclipse has also been reported by \citet{hel00} for EX Hya
during outburst.

\subsection{Optical Spectroscopy}
\begin{figure}
\plotone{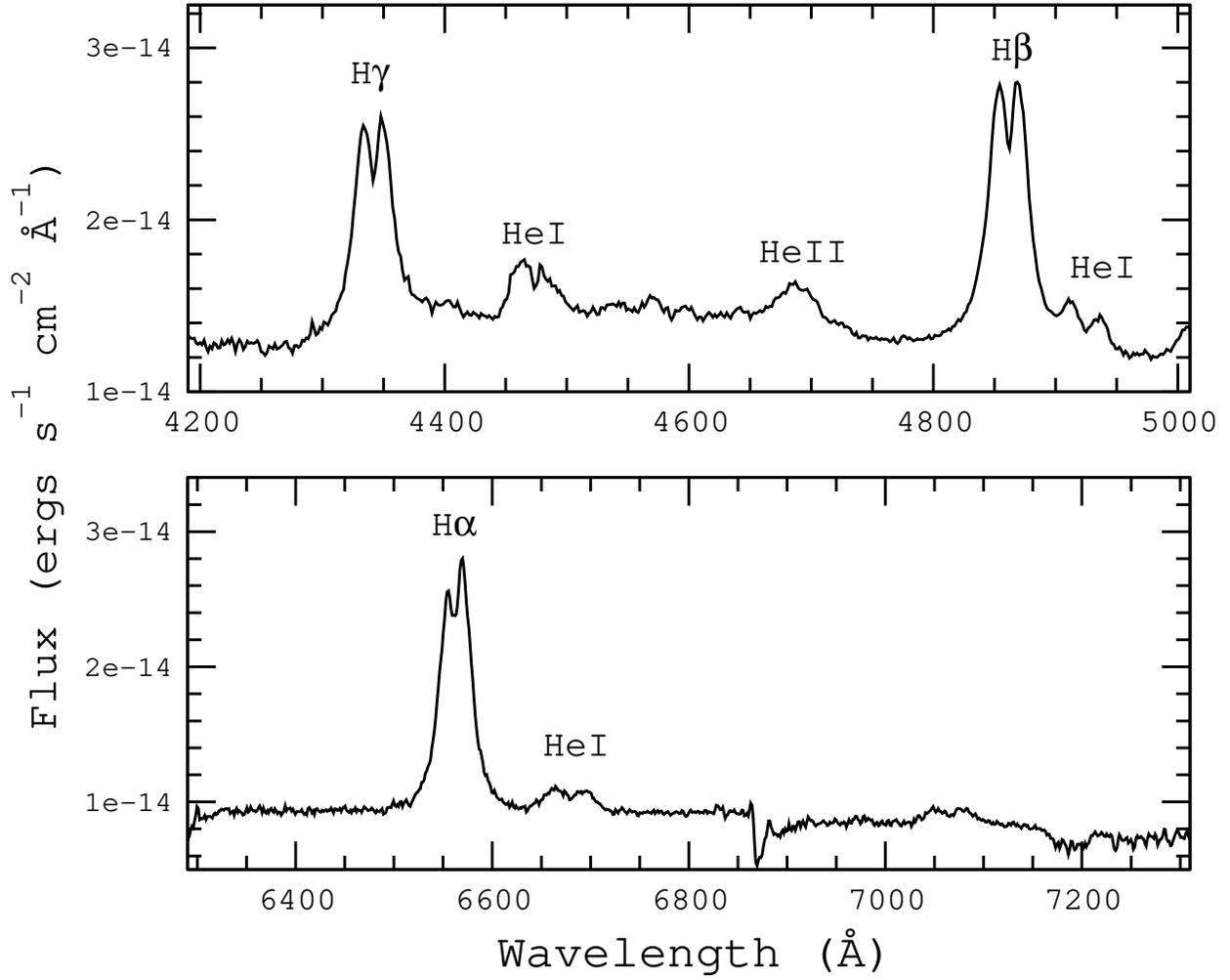}
\caption{Average blue (top) and red (bottom) spectra from
2000 May 28.  \label{avg_spec}}
\end{figure}

\begin{figure}
\plotone{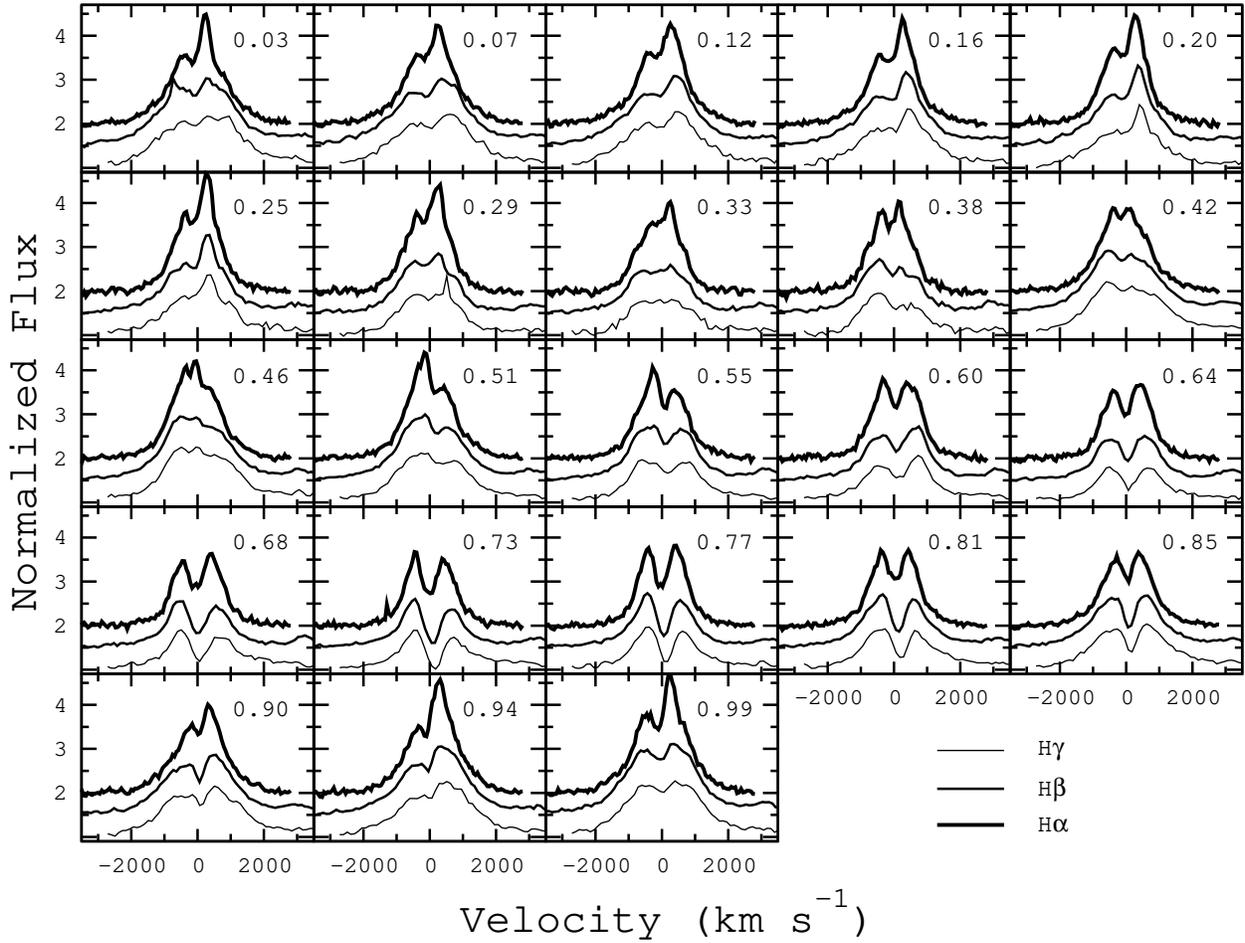}
\caption{$\ha$, $\hb$, and $\hg$ velocity line profiles shown over
the binary orbital period (phase noted in the upper right corner of
each plot) for data from 2000 May 28.  The S-wave component is clearly
visible as a red-shifted or blue-shifted component during most binary
phases.  The $\hb$ and $\ha$ line profiles have been shifted upward in
flux by 0.5 and 1.0, respectively.  \label{hlines}}
\end{figure}

APO spectra from 2000 May 28 are shown in Figures~\ref{avg_spec} and
\ref{hlines}.  Figure~\ref{avg_spec} presents the mean blue and red
spectra, which contain strong Balmer emission lines, as well as
\ion{He}{1} and \ion{He}{2} emission.  Figure~\ref{hlines} presents
velocity line profiles of the $\hg$, $\hb$, and $\ha$ emission lines
plotted over the binary orbital period.  The lines exhibit a
double-peaked shape over all binary phases, indicative of emission
from an accretion disk, and have average FWHM values of $\sim50$\AA,
or $\sim2000-3000\kms$.

\subsubsection{The S-wave Component}\label{swave}
The S-wave component of the emission lines, contributed by the hot
spot on the outer edge of the accretion disk, is clearly visible as a
red-shifted or blue-shifted component in the $\ha$, $\hb$, and $\hg$
emission lines throughout most binary phases.  The exact behavior of
the S-wave component varies between the emission lines.  In $\ha$, the
S-wave is red-shifted during the early binary phases of
$\pb=0.03-0.38$ and has strongest zero-velocity emission during
$\pb=0.4-0.5$, the phases during which we are afforded a view of the
hot spot on the opposite side of the accretion disk.  The S-wave
component is blue-shifted from $\pb=0.51-0.60$, is missing (or at
least much less prominent) from $\pb=0.64-0.85$, the phases during
which we would expect to see the hot spot directly on the near edge of
the disk, and is red-shifted again at $\pb=0.90-0.99$.

In the $\hb$ and $\hg$ emission lines, the S-wave component follows
the general behavior of that seen in the $\ha$ line, with a few
exceptions.  The S-wave component is at zero velocity at $\pb\sim0$
and $\pb\sim0.3$, where the $\ha$ S-wave component is red-shifted.  It
also becomes blue-shifted at earlier phases, around $\pb\sim0.4$ for
both the $\hb$ and $\hg$ lines, while it is still at zero-velocity in
the $\ha$ line.  The S-wave component also appears to bounce back and
forth between being red- and blue-shifted during binary phases
$\pb=0.60-0.85$.  Another intriguing feature of the $\hg$ and $\hb$
emission lines during these phases is that there is little to no
zero-velocity emission.

Taking the general behavior of the S-wave component from all three
lines, we can infer geometric properties of the hot spot and bulge on
the outer edge of the accretion disk.  The S-wave component is
red-shifted from $\pb\sim0.85-0.3$, at zero velocity around
$\pb\sim0.3-0.4$, and is blue-shifted from $\pb\sim0.4-0.6$.  After
$\pb\sim0.6$, the S-wave component disappears.  The zero velocity
emission phase at $\pb\sim0.3-0.4$ tells us that the S-wave component
is dominated by emission on the opposite side of the disk, between
phases $\pb\sim0.8-0.9$ on the accretion disk.  This emission
therefore appears as a red-shifted component from $\pb\sim0.85-0.3$.
The S-wave is only blue-shifted when this component moves toward the
observer, during phases $\pb=0.4-0.6$.  At phase $\pb=0.6$, material
begins to obscure our view of the hotter parts of the hot spot, and so
the S-wave component disappears.  From the missing S-wave emission
during phases $\pb=0.60-0.85$, and the absence of any zero velocity
component in the emission lines, we can infer that there is vertically
extended material along the outer edge of the disk obscuring the
S-wave and zero velocity components starting at $\pb\sim0.6$.  This
extended material is likely irradiated along its inner edge by the
white dwarf, and gives rise to the S-wave emission.

\begin{figure}
\plotone{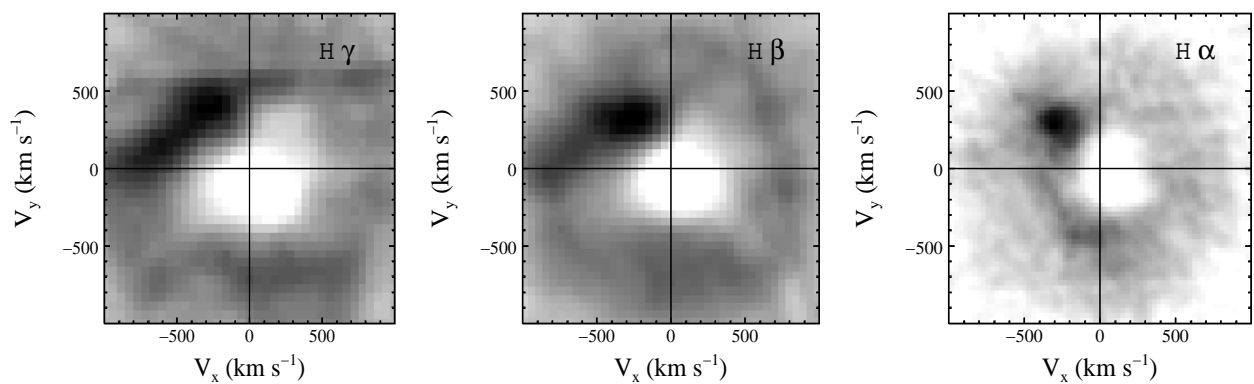}
\caption{From left to right, tomograms over the binary phase for the
$\hg$, $\hb$, and $\ha$ emission lines from spectra obtained during
the night of 2000 May 28.  Each tomogram shows a bright spot at
$\pb\sim0.8$.  The $\hg$ and $\hb$ tomograms show enhanced emission
from the hot spot toward the inner edge of the disk.  Extended
emission (roughly half of the orbital phase) is seen along the outer
edge of the accretion disk in the $\ha$ tomogram.  The phasing of the
tomograms follows standard convention, where binary phase 0 is at 12
o'clock on the plot and phase increases clockwise.  \label{tom}}
\end{figure}

Further evidence for extended emission along the outer edge of the
disk is seen in the Doppler tomogram for the $\ha$ emission line,
displayed in Figure~\ref{tom}.  The tomograms shown in the Figure are
measured from spectra obtained on 2000 May 28 and are plotted over
binary orbital phase.  The $\ha$ tomogram shows enhanced emission
along the outer edge of the accretion disk, for approximately $0.5$ in
phase.  Extended emission along the outer edge of the accretion disk
is also indirectly observed in the EUV data of EX Hya, in the form of
absorption of the EUV emitting region \citep{bel02}.  This absorption
also lasts for approximately half of an orbital phase.

Each tomogram exhibits a bright area of emission at $\pb\sim0.8$ --
this is the hot spot and is the origin of the S-wave component.  The
$\hg$ and $\hb$ tomograms also show enhanced emission that appears to
extend from the region at $\pb\sim0.8$ toward the inner regions of the
disk ($R_{\rm in}\la6.5\rwd$, see \S\ref{rv}).  \citet{hel00} observed
evidence of disk overflow in EX Hya during its 1998 outburst.  In
X-rays, the overflow stream was indirectly observed as an $\wW$ beat
modulation, and in the optical, the overflow stream was detected via
its eclipse and from line emission at the site of the stream impact
with the magnetosphere.  We checked AAVSO data of EX Hya for the time
period immediately preceding and following our observations, but the
available data show no evidence of an outburst.  In fact, the data
points for EX Hya show it sitting at $13-12.5$ mag, which is its
normal quiescent brightness.  This would rule out a period of enhanced
mass transfer as the cause of the overflowing material inferred from
our data, as such an event would likely be indicated by an increase of
the system brightness.

Tomograms plotted over the white dwarf spin phase were also
constructed; however, these show no coherent emission sites.  It has
also been shown previously \citep{hel99} that tomograms of EX Hya
folded on the spin phase reveal little information.

\subsubsection{Radial Velocities}\label{rv}
Measuring the central wavelengths of the H emission lines proved to be
a challenge, as the double-peaked nature of the lines and the
contribution of the S-wave component made it difficult to use a single
Gaussian profile for determining line parameters.  We therefore used a
method for fitting a double-Gaussian profile to the emission lines,
which measures radial velocities from emission line wings
\citep[D. W. Hoard 2004, private communication, for a description of
double-Gaussian fitting see][]{sha83,sha86}.

\begin{figure}
\plotone{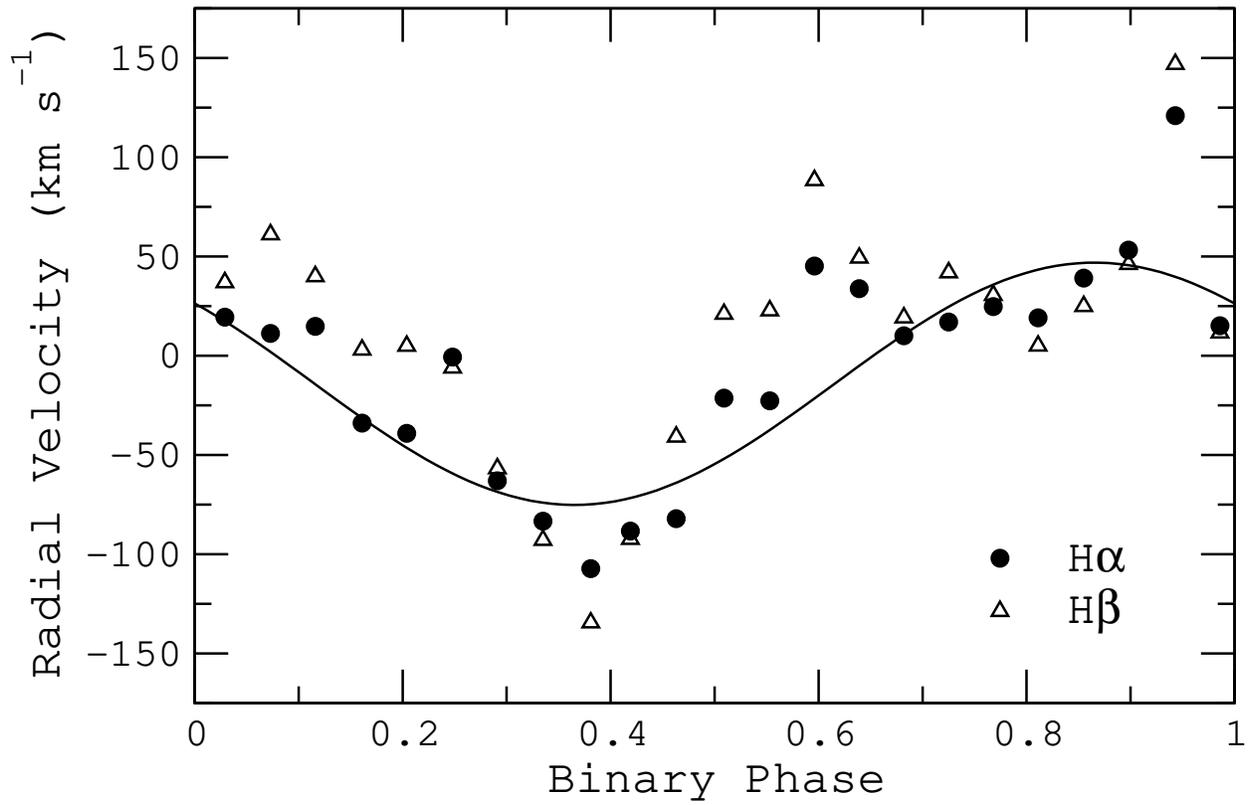}
\caption{Radial velocities of the $\ha$ and $\hb$ optical emission
lines folded on the binary phase. The large negative velocities at
$\pb\sim0.4$ may indicate overflowing material.  The sine curve
represents a fit to the $\ha$ radial velocities. \label{opt_rv}}
\end{figure}

We created radial velocity curves for the $\hg$, $\hb$, and $\ha$
emission lines.  Figure~\ref{opt_rv} displays the radial velocity
curves for the $\ha$ and $\hb$ lines.  Strong asymmetries due to
emission from the hot spot prevented accurate determinations of $K_1$
or $\gamma$ from fitted sine curves, but a fit of the form $v=\gamma +
K_1\sin2\pi(\pb-\phi_0)$ to the $\ha$ line (shown on
Figure~\ref{opt_rv}) gives $\gamma=-14\pm1\kms$, $K_1=61\pm6\kms$, and
$\phi_0=0.115\pm0.016$.  The $K_1$ value of $61\pm6\kms$ is in
agreement with our previous value of $K_1=59.6\pm2.9\kms$ determined
from narrow UV emission lines \citep{bel03}, but slightly lower than a
previous value of $69\pm9\kms$ determined from optical emission lines
by \citet{hel87}.

We also calculated the radial velocity of the S-wave component in each
of the H emission lines.  We found that the velocity amplitude of the
S-wave emission places the emission site at $9\times10^{10}\cm$ from
the center of the white dwarf for the $\ha$ component and at
$3\times10^{10}\cm$ from the center of the white dwarf for the $\hb$
and $\hg$ components.  The smaller value obtained from the $\hb$ and
$\hg$ radial velocities may be due to contamination from the overflowing
material.  If we assume that the S-wave component originates at the
outer edge of the accretion disk, the placement of the $\ha$ S-wave
component gives an outer disk radius of $R_{\rm out}\sim10^{11}\cm$.
The inner accretion disk radius may be measured from the high velocity
emission line wings, which extend at least to $1500\kms$
(Figure~\ref{hlines}).  This velocity gives an upper limit to the inner
disk radius of $R_{\rm in}=6.5\rwd=5\times10^{9}\cm$ for a $0.8\msun$
white dwarf or $R_{\rm in}=2.9\rwd=3\times10^{9}\cm$ for a $0.5\msun$
white dwarf.

\section{COMPARISON OF DATA AT ALL WAVELENGTHS}\label{compare}
A comparison of data from all wavelength regimes can lead us to an
overall picture of EX Hya.  Signatures of an extended bulge along the
outer edge of the accretion disk are seen in the EUV photometry as
absorption of the EUV emitting region on the white dwarf surface, in
the optical spectroscopy as enhanced emission in the $\ha$ emission
line tomogram, and as absorption of the zero velocity components of
the $\hb$ and $\hg$ emission lines during certain binary phases.  The
absorption at phases $\pb=0.55-1.1$ in the EUV photometry match well
with the absorption in the $\hb$ and $\hg$ emission lines, seen at
$\pb=0.6-0.9$.  The fact that the bulge is not present as a source of
continuum emission in the binary phased light curves implies that the
bulge material is optically thin.

Some form of overflowing material is seen in the Doppler tomograms of
the $\hg$ and $\hb$ emission lines as enhanced emission extending from
the hot spot toward the inner edge of the accretion disk.  Other
evidence for overflowing material comes from the blue-shifted
velocities of the Balmer lines at $\pb\sim0.4$.  One may also infer
the existence of an extended/overflowing hot spot from the extended
egress shoulder of the binary eclipse during spin minimum phases,
which implies that an additional source, other than the white dwarf,
is coming out of eclipse.

Some inconsistencies remain.  The first is the phase of spin maximum
as derived from light curves of continuum and emission line fluxes.
X-ray spin maximum occurs at $\ps\sim0$, EUV at $\ps=0.115\pm0.001$,
UV continuum at $\ps=0.01\pm0.05$, UV emission lines at
$\ps=(0.05-0.08)\pm0.05$, and optical at $\ps=0.12\pm0.02$.  The
difference in the phase of spin maximum has been reported previously
by \citet{hel00} for their simultaneous X-ray and optical
observations.  An extended accretion region on the white dwarf surface
would be an initial suggestion for the difference in phase of spin
maximum; however, one would expect that the high energy spin maxima
would be coincident, but this is not the case.

Another inconsistency is the phase of the binary eclipse across
wavelength regimes.  X-ray and optical data show the binary eclipse at
$\pb\approx0.0$, while EUV data displays two eclipses near, but not
at, $\pb=0$: one at $\pb=0.97$ and the other at $\pb=1.04$.  The EUV
eclipse at $\pb=1.04$ could be an eclipse of the disk overflow stream
impact with the magnetosphere, though this region would not be
expected to be bright in the EUV.  But there still remains the phase
difference of the binary eclipse.  The difference between the
EUV and X-ray eclipses could be reconciled by the fact that the {\it
RXTE} coverage of those phase bins is not nearly as extensive as the
{\it EUVE} coverage.  Perhaps observations that provide better phase
sampling would show the X-ray eclipse slightly earlier in phase.  In
such a case, then, a physical displacement between the higher and
lower energy emitting regions could be the cause of the phase
difference of the binary eclipses.

\section{CONCLUSIONS}
We have presented simultaneous X-ray and optical photometry and
optical spectroscopy of the IP EX Hya obtained as part of a large
multiwavelength observational campaign of EX Hya.  The data provide
evidence for an extended bulge along the outer edge of the accretion
disk, corroborating our EUV results \citep{bel02}.  Over the binary
orbital period, the zero-velocity component of the optical emission
lines is seen to be absorbed during binary phases $\pb=0.6-0.85$.  The
$\ha$ tomogram also shows enhanced emission along the outer edge of
the accretion disk at phases $\pb=0.6-0.85$.  The $\hb$ and $\hg$
tomograms also indicate that some amount of material may be extended
from the hot spot toward the inner regions of the accretion disk.

Combined together, the data suggest that EX Hya may be experiencing a
period of enhanced mass transfer, as the data presented here bear
similarities to the outburst data of 1998.  The system remained,
however, at its quiescent brightness of $\sim13$ mag during the
observations.  Perhaps there was a period of enhanced mass transfer
that was not recorded, or EX Hya may not have returned entirely to its
quiescent state after outburst.  Finally, we would like to call the
attention of the reader to an initial light curve of EX Hya published
by \citet[][Fig. 5]{mum67} prior to knowledge of EX Hya as an IP
(i.e., no spin modulation was subtracted from the light curve).  This
light curve (which has $\Delta{\rm m}\sim0.4$ between eclipses) shows
none of the modulation associated with the white dwarf spin period
that is so prevalent in our 2000 photometric data ($\Delta{\rm
m}\sim0.7$ between eclipses).  Obviously, conditions within EX Hya,
e.g., a larger bulge or extended accretion curtains, have changed in
order to produce the enhanced emission.  It is apparent that obtaining
over one million seconds of data has only increased the number of
questions we have about EX Hya.

\acknowledgments 
KEB thanks P. Bradley for very helpful discussions and D. W. Hoard for
providing his double-Gaussian fitting routines.  We thank Dr. Jean
Swank, the RXTE project scientist, for the allocation of the Target of
Opportunity observing time, and the RXTE Science Operations Center for
scheduling these observations.

\newpage

\clearpage
\newpage

\begin{deluxetable}{lll}
\tablecolumns{3}
\tablewidth{0pc}
\tablecaption{Data Obtained for EX Hya during the 2000 May -- June 
Observational Campaign.\label{obslog}}
\tablehead{
\colhead{Instrument} & \colhead{Observations\tablenotemark{a}} 
& \colhead{Bandpass}
}
\startdata
Chandra    &         60 ks  spectroscopy  & $1-20$ \AA \\
RXTE*       &        15 ks photometry and spectroscopy, & $1-5$ \AA \\
            &       2000 May 18, 30 &   \\
USA         &        41.2 ks  photometry & $1-10$ \AA  \\
EUVE       &          1000 ks  photometry and spectroscopy & $70-180$ \AA \\
FUSE       &          28 ks spectroscopy & $800-1200$ \AA \\
HST        &          6 orbits spectroscopy & $1100-1700$ \AA \\
APO*         &   2 nights spectroscopy, & $4200-5000$, $6300-7300$ \AA  \\
         &  2000 May 15 \& 28  & \\
BO*          &       16 nights photometry, & $B, V, R$\\
       & 2000 May 1--5, 7, 9, 10, 12--14, 25, 26 \& Jun 2, 3, 7 & \\
RHO*         &       10 nights photometry, & $I$  \\
            & 2000 May 5, 8, 12, 18--20, 26--28, 31 & \\
UKIRT       &        1 night spectroscopy & $1.9-2.5\,\micron$  \\
\enddata
\tablenotetext{a}{Times given are in UT.}
\tablenotetext{*}{Data presented in this paper.}
\end{deluxetable}

\begin{deluxetable}{ccc}
\tablecolumns{3}
\tablewidth{0pc}
\tablecaption{Photometric frequencies appearing in the X-ray and optical 
photometry.\label{freq}}
\tablehead{
\colhead{Observed frequency} & \colhead{Frequency} & \colhead{Period}\\
\colhead{} & \colhead{(day$^{-1}$)} & \colhead{(min)} }
\startdata
$\w$         & 21.48 & 67.03  \\
$\w/2$       & 10.74 & 134.05 \\
$\W$         & 14.66 & 98.26  \\
$\W/2$       & 7.33  & 196.51 \\
$\wW$        & 6.83  & 210.88 \\
$\wWW$       & 7.83  & 183.98 \\
\enddata
\end{deluxetable}

\begin{deluxetable}{cccc}
\tablecolumns{4}
\tablewidth{0pc}
\tablecaption{$\Theta$ values of frequencies seen in the optical and X-ray 
thetagrams.\label{theta}}
\tablehead{
\colhead{} & \multicolumn{3}{c}{$\Theta$\tablenotemark{a}} \\
\cline{2-4}
\colhead{Frequency}& \colhead{$B$}  &  \colhead{$I$}  & \colhead{X-ray} }
\startdata
$\w$         & 0.42 & 0.56 & 0.76 \\
$\W$         & 0.92 & 0.80 & 0.89 \\
$\wW$        & 0.55 & 0.64 & \nodata \\
$\wWW$       & 0.59 & 0.62 & \nodata \\
\enddata
\tablenotetext{a}{95\% confidence level is at $\Theta=$ 0.97 ($B$),
$>0.99$ ($I$), 0.96 (X-ray), 99\% confidence level is at $\Theta=$ 0.95
($B$), $>0.99$ ($I$), 0.93 (X-ray)}
\end{deluxetable}

\begin{deluxetable}{cccc}
\tablecolumns{4}
\tablewidth{0pc}
\tablecaption{Sinusoidal fits to the spin phased photometry.
\label{sp_fit}}
\tablehead{
\colhead{Bandpass} & \colhead{$A$\tablenotemark{a}} & 
\colhead{$B$\tablenotemark{a}} & \colhead{$\phi_0$}}
\startdata
$B$ &  	$-0.75\pm0.01$ & $-0.20\pm0.01$ & $0.87\pm0.02$ \\
$V$ &	$-0.27\pm0.01$ & $-0.17\pm0.01$ & $0.87\pm0.02$ \\
$R$ & 	$-0.13\pm0.01$ & $-0.17\pm0.01$ & $0.87\pm0.02$ \\
$I$ & 	$0.02\pm0.01$  & $0.13\pm0.01$  & $0.82\pm0.02$ \\
X-ray & $10.29\pm0.01$ & $1.09\pm0.15$  & $0.74\pm0.02$ \\
\enddata
\tablenotetext{a}{$A$ and $B$ have units of $\Delta$m for $BVRI$ and
counts sec$^{-1}$ PCU$^{-1}$ for X-ray.}
\end{deluxetable}


\begin{thebibliography}{99}
%
\bibitem[Allan et al.(1998)]{allan98}
Allan, A., Hellier, C., \& Beardmore, A.
1998, \mnras, 295, 167
%
\bibitem[Belle et al.(2002)]{bel02}
Belle, K. E., Howell, S. B., Sirk, M. M., \& Huber, M. E.
2002, \apj, 577, 359
%
\bibitem[Belle et al.(2003)]{bel03}
Belle, K. E., Howell, S. B., Sion, E. M., Long, K. S., \& Szkody, P.
2003, \apj, 587, 373
%
\bibitem[Beuermann \& Osborne(1988)]{beuer88}
Beuermann, K. \& Osborne, J. P.
1988, \aap, 189, 128
%
\bibitem[Cordova et al.(1985)]{cord85}
C\'{o}rdova, F. A., Mason, K. O., \& Kahn, S. M.
1985, \mnras, 212, 447
%
\bibitem[Fujimoto \& Ishida(1997)]{fuj97}
Fujimoto, R. \& Ishida, M.
1997, ApJ, 474, 774
%
\bibitem[Hellier(1999)]{hel99}
Hellier, C.
1999, \apj, 519, 324
%
\bibitem[Hellier et al.(2000)]{hel00}
Hellier, C., Kemp, J., Naylor, T., Bateson, F. M., Jones, A., 
Overbeek, D., Stubbings, R., \& Mukai, K.
2000, MNRAS, 313, 703
%
\bibitem[Hellier et al.(1987)]{hel87}
Hellier, C., Mason, K. O., Rosen, S. R., \& C\'{o}rdova, F. A.
1987, \mnras, 228, 463
%
\bibitem[Hellier \& Sproats(1992)]{ephem}
Hellier, C. \& Sproats, L. N.
1992, IBVS, 3724
%
\bibitem[Hoogerwerf et al.(2004)]{hoo04}
Hoogerwerf, R., Brickhouse, N. S., \& Mauche, C. W.
2004, \apj, 610, 411 (astro-ph/0403665)
%
\bibitem[Mauche et al.(2001)]{mau01}
Mauche, C. W., Liedahl, D. A., \& Fournier, K. B.
2001, \apj, 560, 992
%
\bibitem[Mauche et al.(2003)]{mau03}
Mauche, C. W., Liedahl, D. A., \& Fournier, K. B.
2003, \apj, 588, L101
%
\bibitem[Mumford(1967)]{mum67}
Mumford, G. S.
1967, ApJS, 15, 1
%
\bibitem[Mukai et al.(2003)]{muk03}
Mukai, K., Kinkhabwala, A., Peterson, J. R., Kahn, S. M., \& Paerels, F.
2003, \apj, 586, L77
%
\bibitem[Rosen et al.(1988)]{rosen88}
Rosen, S. R., Mason, K. O., \& C\'{o}rdova, F. A.
1988, \mnras, 231, 549
%
\bibitem[Shafter(1983)]{sha83}
Shafter, A. W.
1983, \apj, 267, 222
%
\bibitem[Shafter et al.(1986)]{sha86}
Shafter, A. W., Szkody, P., \& Thorstensen, J. R.
1986, \apj, 308, 765
%
\bibitem[Siegel et al.(1989)]{siegel89}
Siegel, N., Reinsch, K., Beuermann, K., van der Woerd, H., \&
Wolff, E.
1989, \aap, 225, 97
%
\bibitem[Stellingwerf(1978)]{stell78}
Stellingwerf, R. F.
1978, \apj, 224, 953
%
\end{thebibliography}
\end{document}